# A vision for a colorectal digital twin that enables proactive and personalized disease management

**Sayed Chhattan Shah¹, Andrea Townsend-Nicholson², Spiros Denaxas², Pablo Lamata³ and Manish Chand⁴**

¹ Southern Illinois University, Carbondale, IL, USA

² Structural & Molecular Biology, Division of Biosciences, University College London, Gower Street, London, UK

³ King's College London, London, UK

⁴ University College London Hospitals NHS Foundation Trust, London, UK

*Abstract —* **Colorectal cancer, inflammatory bowel disease, and diverticular disease are progressive conditions that affect millions of individuals worldwide and impose substantial clinical and economic burdens. Early detection and personalized management are essential for slowing disease progression and improving patient outcomes. Current care pathways rely primarily on episodic clinical encounters, laboratory testing, and reactive interventions, limiting early detection and personalized longitudinal management. This paper introduces a conceptual framework for an integrated colorectal digital twin that supports non-invasive, continuous monitoring and personalized disease management. The framework integrates multimodal physiological and behavioral data streams, hybrid mechanistic–machine learning modeling of colorectal function, and a personalized artificial intelligence engine to support proactive disease management. Rather than presenting a deployed clinical system, this work outlines a clear vision and a structured approach for colorectal digital twins, identifying key technical, modeling, and translational challenges necessary for future implementation and validation.**

## ■ INTRODUCTION

The intestine, an integral part of the digestive system, consists of the small intestine and the large intestine. The large intestine includes the colon and rectum [1]. The colon, being the longest part, receives digested food and passes waste to the rectum, which is the lower part of the large intestine. The rectum receives waste from the colon and stores it until it is released through the anus [2]. Cancer is a group of diseases involving abnormal cell growth with the potential to invade or spread to other parts of the body. Colorectal cancer, also called colon cancer, rectal cancer, or bowel cancer, starts in the colon or rectum and is the third most common cancer in the world [1]. Inflammatory bowel disease (IBD) refers to a group of chronic, progressive diseases that cause inflammation in the intestines. The two main types of IBD are ulcerative colitis and Crohn's disease. Diverticular disease is another prevalent condition that results in the formation of small pouches in the walls of the large intestine. Additionally, several other conditions, including rectal ulcers, hemorrhoids, irritable bowel syndrome, and large bowel obstruction, are associated with the large intestine [3, 4].

In the UK, colorectal cancer remains a significant health concern with over 42,000 new cases reported annually, leading to more than 16,500 deaths each year [5]. Currently, there are approximately 268,000 individuals living with colorectal cancer across the country, alongside more than 500,000 individuals with IBD [5]. The UK allocates 1.7 billion GBP solely for the care of colorectal cancer patients. Additionally, millions of pounds are spent on the treatment of IBD and diverticular disease. The annual cost to employers due to IBD is estimated at around 600 million GBP.

The US reported 142,462 new cases of colorectal cancer in 2019 and over 51,000 fatalities [6, 7]. The healthcare costs in the country have reached a massive 4.5 trillion USD in 2021, with an average spending of 12,914 USD per person, making up 18.3% of the GDP [6]. Specifically, cancer care expenses amounted to 200 billion USD in 2020, of which 24 billion USD were allocated for colorectal cancer treatment, ranking it second in treatment expenses among all cancer types [7, 8]. Around 2.8 million Americans live with IBD, with total annual healthcare costs estimated at 8.5 billion USD in 2018. Additionally, millions of dollars are allocated towards addressing various diseases of the large intestine, such as rectal ulcers, hemorrhoids, anal fissures, large bowel obstruction, and irritable bowel syndrome, collectively affecting millions of individuals.

Colorectal cancer and IBD are progressive conditions, and early detection is crucial for slowing disease progression and reducing pain and discomfort. With timely diagnosis and appropriate interventions, such as medical treatments, lifestyle modifications, stress reduction strategies, and regular follow-ups, patients can experience significant improvements in both quality of life and clinical outcomes.

Current clinical practice relies primarily on in-person consultations, episodic laboratory testing, and patient self-reporting. While these methods are clinically validated, primary care physicians often face substantial workload pressures and limited time per patient, and dependence on in-person visits and laboratory testing is time-consuming and costly for both patients and healthcare systems.

Existing approaches to colorectal disease management predominantly rely on single-modality monitoring technologies, isolated prediction models, or reactive management tools. To date, there is no colorectal-specific digital twin framework that integrates multimodal non-invasive data, physiologically grounded colorectal modeling, and personalized artificial intelligence for continuous, patient-specific monitoring and proactive disease management.

This paper posits that emerging digital health technologies create an opportunity to rethink how colorectal health is conceptualized and monitored. Rather than focusing on isolated measurements or episodic assessments, colorectal health may be better represented as a continuously evolving physiological system. Colorectal Twin is introduced as a conceptual framework that advances this perspective. This work contributes to the emerging digital twin literature by:

- Introducing the first colorectal-specific digital twin framework

- Combining mechanistic colorectal physiology with adaptive machine learning in a hybrid framework

- Extending retrieval-augmented large language models into longitudinal colorectal disease management

- Proposing a staged translational validation pathway grounded in gastroenterology practice

**BACKGROUND AND RELATED WORK**

**Healthcare Digital Twins**

A human digital twin is a virtual representation of a human or a specific organ system designed to simulate physiological behavior, monitor health, and support predictive and prescriptive decision-making. Digital twins have been explored for several organ systems, including the liver [9], the heart [10], and the lumbar spine [11]. These efforts demonstrate the feasibility of physiology-based digital twins. However, they also highlight a notable gap: the absence of colorectal-specific digital twin frameworks that account for the unique complexity, heterogeneity, and often silent progression of colorectal diseases.

### Data Collection

Colorectal health monitoring remains dominated by fragmented data sources and modality-specific tools rather than integrated, system-level representations. Most remote monitoring approaches rely either on self-reported questionnaires or wearable and ambient sensing technologies [12–21]. While these technologies provide valuable physiological signals, questionnaires fail to capture objective biological measurements, and wearable or ambient sensors typically generate indirect or isolated data streams. Consequently, existing data collection approaches do not provide a comprehensive, physiologically grounded representation of colorectal health, nor do they enable continuous, integrated assessment of colorectal function and disease progression.

Representative examples include:

- Oura Ring (Oura Health, Finland), measures blood oxygen levels, heart rate and heart rate variability, respiration rate and temperature variations. It also tracks movement and activity and can be used to determine inflammation levels.

- Google Pixel Watch (Google LLC, USA), tracks heart rate and breathing rate, skin temperature measurement, blood oxygen tracking, fall detection, and stress management

- Accelerometer (PLUX Biosignals, Portugal), used for activity and movement monitoring

- cardioBAN (PLUX Biosignals, Portugal), a wearable system that captures raw ECG and motion data.

- Corti (EnLiSense LLC, USA), which monitors stress, sleep, and metabolic health

- IBD Aware (EnLiSense LLC, USA), a wearable system that monitors IBD-related biomarkers such as Calprotectin, C-Reactive Protein (CRP), IL-6, and TNF-α

- Electrogastrography (EGG) sensor (PLUX Biosignals, Portugal), designed to record gastric electrical activity in a non-invasive manner

- Video sensor or smart toilet, developed to detect selected disease markers in stool and urine.

### Imaging-based Disease Assessment

Numerous AI-based approaches have been explored for colorectal disease assessment, using models such as convolutional neural networks (CNNs), recurrent neural networks (RNNs), and gradient-boosting methods [25-27]. Across colorectal cancer (CRC), inflammatory bowel disease (IBD), and diverticular disease, these studies demonstrate the promise of AI for diagnosis and assessment. However, many existing models are trained on limited input modalities, such as imaging or structured clinical records, or are relatively limited datasets, which may constrain their generalizability and broader clinical applicability. For example, studies in [22–23] applied convolutional neural networks to endoscopic or histopathological images to assess colorectal abnormalities, while [24] employed a Scaled-YOLOv4 framework for lesion detection in colorectal imaging data. These systems show strong performance in controlled procedural settings but typically do not incorporate longitudinal physiological signals or multimodal patient data.

### EHR-based Disease Assessment

Beyond imaging, machine learning models have been developed using electronic health records (EHRs) to predict colorectal cancer risk [25–28]. For instance, long short-term memory (LSTM) networks and gated recurrent units (GRUs) have been used to model patterns in electronic medical records for CRC prediction [25]. Random forest and neural network models to estimate CRC risk among adults aged 35–50 using structured EHR variables [26]. Similarly, machine learning techniques have been used to predict early-onset colorectal cancer in individuals younger than the recommended screening age by analyzing structured EHR data [27], and a logistic regression model trained on multimodal EHR features has been developed to estimate CRC risk [28].

## LLM-based Disease Management

Large language models (LLMs), such as ChatGPT, have increasingly been explored in healthcare applications, including clinical decision support, transplant management, and dietary guidance for gastrointestinal conditions. However, multiple studies have reported that LLM performance can be unreliable in clinical contexts due to hallucinations and factual inaccuracies, raising safety concerns for high-risk diseases such as colorectal cancer [29–31].

To mitigate these limitations, recent research has investigated the integration of Retrieval-Augmented Generation (RAG) with LLMs to ground responses in curated medical knowledge [32–35]. For example, LiVersa [32] combines LLMs with RAG to support hepatitis B management. ALMANAC [33] demonstrated improved performance over baseline LLMs in clinical reasoning tasks, and disease-specific RAG pipelines have incorporated structured guidelines into LLM-based workflows [34]. Although these approaches report improved accuracy in controlled evaluations, they remain largely guideline-driven and text-based. They do not incorporate patient-specific physiological signals, longitudinal biomarker data, or individualized disease trajectory modeling.

To the best of our knowledge, there is currently no published system that integrates a RAG–enhanced LLM for continuous colorectal disease management.

The management of colorectal diseases is inherently complex. CRC, IBD, and diverticular disease are frequently influenced by comorbid conditions such as metabolic disorders, cardiovascular disease, and lifestyle-related factors. Effective management therefore requires coordinated, patient-specific strategies that may include continuous symptom monitoring, inflammatory marker tracking, medication adherence, dietary planning, physical activity, weight management, stress regulation, and psychosocial support [2, 3, 8]. Current LLM-based systems do not capture these dynamic physiological interdependencies.

## COLORECTAL DIGITAL TWIN OVERVIEW

The Colorectal Twin is envisioned as an integrated, conceptual framework for non-invasive monitoring, early detection, and personalized management of colorectal diseases. The system architecture presented here is intentionally high-level and implementation-independent, designed to establish research direction and guiding design principles. Colorectal Twin is structured around three tightly coupled layers: (1) a multimodal data integration layer, (2) a hybrid colorectal digital twin model, and (3) a personalized AI engine.

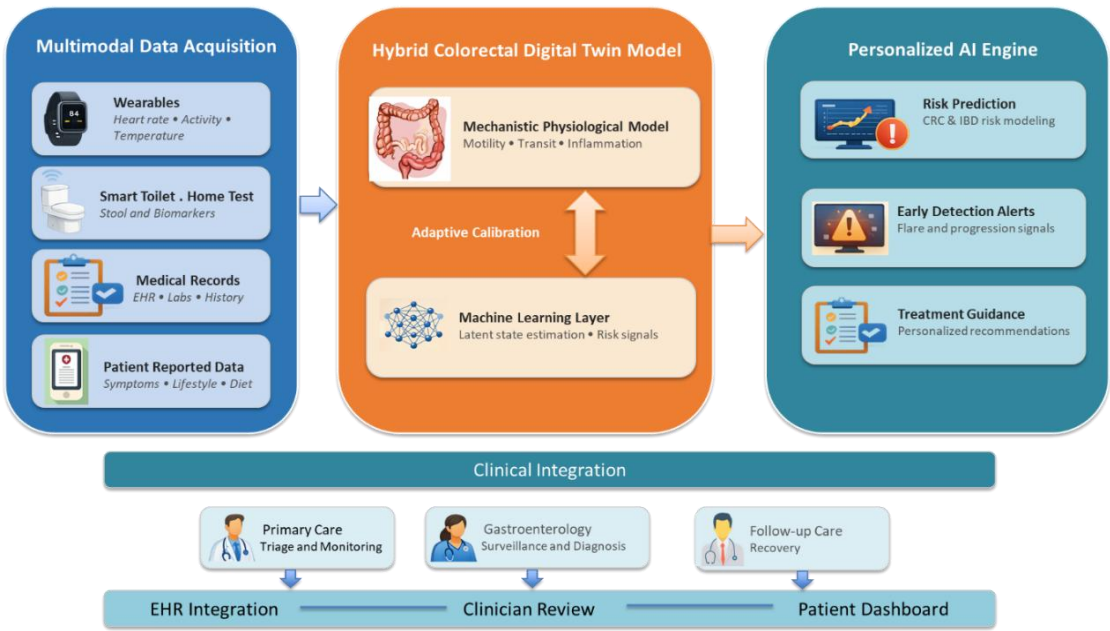

Figure 1: Conceptual Framework for Colorectal Twin

**Multimodal Data Integration**

Colorectal diseases are associated with a wide range of symptoms, including abdominal pain, blood in the stool, and fatigue, and are influenced by risk factors such as age, race, smoking, diabetes, family history, diet, and physical inactivity [4]. Table 1 summarizes common symptoms and risk factors associated with colorectal cancer (CRC), inflammatory bowel disease (IBD), and diverticular disease (DD).

Table 1: Common symptoms and risk factors associated with colorectal cancer, inflammatory bowel disease, and diverticular disease.

|  | CRC | IBD | DD |
|---|---|---|---|
| Abdominal pain |  | x |  |
| Lower abdomen pain |  |  | x |
| Frequent gas pains | x |  |  |
| Bloating | x |  | x |
| Diarrhea | x | x | x |
| Constipation | x |  | x |
| Blood in the stool | x | x |  |
| Mucus in the stool |  | x |  |
| Bleeding from the bottom |  | x |  |
| A change in bowel habits | x |  |  |
| A change in bowel shape | x |  |  |
| Weight loss | x | x |  |
| Cramps | x |  | x |
| Fever |  |  | x |
| Fatigue | x | x |  |
| Nausea |  |  | x |
| Vomiting |  |  | x |
| Tenesmus | x |  |  |

To enable comprehensive and non-invasive monitoring, the multimodal data integration layer is designed to collect and harmonize data from diverse colorectal-related sources, subject to availability, patient consent, and future technical realization. Collectively, these data streams provide a continuous and multidimensional view of colorectal health that cannot be achieved through any single modality alone.

- *Physiological data from wearable sensors:* heart rate, respiratory rate, skin temperature, blood oxygen saturation, activity levels, and stress-related biomarkers

- *Smartphone-based stool phenotyping and fecal biomarker analysis:* AI-driven classification of stool form and color, detection of visible blood or mucus, and camera-assisted quantification of fecal biomarkers such as calprotectin

- *Behavioral and lifestyle metrics obtained from smartphones and wearables:* physical activity patterns, mobility trends, sleep duration and quality, and dietary-related inputs where available

- *Electronic health records and laboratory data:* colonoscopy findings, histopathology reports, fecal immunochemical test results, inflammatory markers, prior diagnoses of CRC, IBD, or DD, comorbidities, and medication history

- *Patient-reported outcomes and symptom questionnaires:* abdominal pain, changes in bowel habits, bloating, fatigue, and other gastrointestinal symptoms

Examples of non-invasive monitoring technologies relevant to these data streams are discussed in the Data Collection section.

This multimodal integration layer forms the foundation of the colorectal digital twin by ensuring that diverse physiological, biochemical, behavioral, and clinical signals are consistently captured, harmonized, and made available for real-time modeling and inference.

**Hybrid Colorectal Digital Twin Model**

A model is a simplified representation of reality, and the scope of colorectal models spans colonic motility, intestinal transit dynamics, mucosal inflammation, epithelial barrier integrity, and interactions between luminal contents and the colonic wall.

Models at the tissue biomechanical level characterize how colon and rectal tissues deform, bear load, and transduce mechanical stimuli. Current models integrate the multilayered structure of the colorectum, nonlinear anisotropic material behavior, and the mechanosensitive neural pathways responsible for encoding stretch and distension. These models provide essential insights into visceral pain, afferent nerve activation, and pathological states such as obstructive disorders [42].

Full-organ finite element models extend mechanistic understanding to the entire colon, enabling simulations of deformation, stress propagation, and interaction with medical devices such as colonoscopes [43]. These approaches expand the capacity to evaluate device safety, predict tissue loading, and explore clinically relevant scenarios that are difficult to study experimentally.

Beyond physical mechanics, models also address biological and pathophysiological processes. The compuGUT framework, for example, simulates interactions between the intestinal microbiota, nutrient flows, and environmental conditions, providing a platform to test hypotheses about diet–flora dynamics and gut ecology in a controlled in silico environment [44]. Meanwhile, quantitative systems pharmacology models of inflammatory bowel disease (IBD) mechanistically capture immune activation, inflammatory signaling, and drug effects, offering a mechanistic foundation for interpreting disease activity. Recently, these models have been paired with machine learning approaches to translate simulated inflammatory markers into clinical disease scores, addressing limitations in biopsy availability and subjective assessment criteria [45].

Mechanistic components of a colorectal digital twin may be implemented using reduced-order compartmental models of intestinal transit [39,40], ordinary differential equation (ODE) systems representing inflammatory dynamics [41], and simplified fluid absorption models. Such models have been widely used in physiological systems modeling and provide a tractable way to capture causal relationships between motility, inflammation, and mucosal function.

In parallel, data-driven machine learning components can be incorporated to estimate latent physiological parameters, correct model mismatch, and capture nonlinear multimodal interactions that are difficult to formalize analytically. Model parameters and states can be updated using Bayesian inference or sequential state estimation techniques such as Kalman filtering, enabling the digital twin to adapt continuously as new patient data become available.

**Personalized AI Engine**

The personalized AI engine builds upon outputs of the hybrid colorectal digital twin by integrating population-level risk models with individualized physiological trajectories. Population-based datasets, including PLCO [36] and UK Biobank [37], provide statistical priors and disease progression patterns, while the patient-specific digital twin supplies continuously updated physiological state estimates. By combining these perspectives, the AI engine can detect deviations from expected disease trajectories, estimate short- and long-term risk probabilities, evaluate treatment response, and generate personalized recommendations. Importantly, the system is designed as a clinician-support tool rather than an autonomous decision-maker, with high-risk alerts requiring human oversight to ensure safety and regulatory compliance.

**RESEARCH CHALLENGES**

Developing a colorectal digital twin presents a set of open scientific, technical, and infrastructural challenges that extend beyond conventional machine learning or remote monitoring systems. These challenges arise from the need to integrate heterogeneous data streams, construct physiologically meaningful and adaptive models, infer latent colorectal states non-invasively, and support continuous operation within real-world healthcare environments. Addressing these challenges represents a multi-year research agenda rather than a single system implementation.

**Multimodal Data Collection and Integration**

Creating an accurate colorectal digital twin requires synthesizing complex, heterogeneous, multimodal data from sources such as:

- wearable sensors (heart rate, respiration, activity, temperature)

- smartphone-based measurements (stool form, color, visible blood or mucus)

- home diagnostic kits (fecal biomarkers)

- electronic health records (EHR), including comorbidities and medications

- patient-reported symptoms and lifestyle data

Because these data originate from diverse devices and platforms with varying communication protocols, sampling frequencies, and data formats, several challenges emerge.

Key Data Challenges

- How can data associated with the colon and rectum be collected in real time using non-invasive methods?

- How can devices equipped with heterogeneous communication technologies be efficiently discovered, synchronized, and managed?

- How can diverse structured and unstructured data streams be standardized, harmonized, and temporally aligned?

- What interoperable interfaces can support integration across sensors, mobile applications, and healthcare information systems?

- How should data be securely stored, filtered, and preprocessed before transmission to higher-level digital twin modeling layers?

- How can limitations of current sensor technologies be mitigated to ensure sufficient physiological coverage of colorectal processes?

- What strategies ensure data completeness and quality sufficient for accurate digital twin calibration?

**Modeling Complexity**

Colorectal function emerges from complex interactions among colonic motility and transit dynamics, mucosal immune activity, fluid absorption processes, and bidirectional communication with systemic physiological and behavioral factors. These processes vary substantially across individuals depending on age, genetic predisposition, diet, microbiome composition, physical activity, stress, medication use, and environmental exposures.

Building a colorectal digital twin therefore requires integrating mechanistic representations of intestinal physiology with data-driven machine learning models capable of capturing nonlinear relationships across multimodal biomarkers, symptom patterns, behavioral signals, and imaging-derived stool features. This integration introduces several core modeling challenges.

Key Modelling Challenges

- How can colonic motility, transit dynamics, mucosal inflammation, and luminal–wall interactions be represented using physiologically meaningful equations while adapting to patient-specific data?

- How can mechanistic physiological models be stably integrated with machine-learning components within a hybrid framework that remains interpretable and computationally tractable?

- How can latent physiological states, such as inflammatory burden, mucosal stability, motility efficiency, and early neoplastic transformation risk, be inferred non-invasively from longitudinal real-world data?

- How can the colorectal digital twin interact with, or be extended to, other organ-level digital twins such as liver, to represent system-wide interactions influencing disease progression?

**Resource Intensive Infrastructure**

A colorectal digital twin requires multiple interconnected layers, including data acquisition, preprocessing and communication, storage, physiological modeling, prediction, and visualization. These layers depend on distributed computing resources and reliable, low-latency communication. Because the system must process continuous multimodal data streams and support real-time inference, significant infrastructural challenges arise.

Key Infrastructure Challenges

- How can diverse processing, communication, and storage requirements be met while ensuring low latency, scalability, and reliability?

- Where should digital twin components be deployed (edge, cloud, hybrid) to optimize performance, privacy, and cost?

- How can microservices architectures and containerization support modular, scalable deployment?

- What strategies can minimize energy consumption in distributed computing environments?

- How can communication between digital twin components remain secure, resilient, and compliant with healthcare data protection standards?

## VALIDATION AND EVALUATION STRATEGY

Although Colorectal Twin is presented as a conceptual framework, its clinical relevance depends on rigorous validation grounded in colorectal pathophysiology and established gastroenterology practice. Validation must therefore assess not only predictive performance, but also the physiological plausibility and clinical actionability of inferred colorectal states.

At the data acquisition level, validation must examine the reliability of colorectal-relevant signals. For example, smartphone-based stool phenotyping should be evaluated against established stool classification systems such as Bristol Stool Form Scale [38]. Non-invasive fecal biomarker measurements, such as calprotectin or fecal immunochemical test (FIT) signals, should be compared with laboratory-based analyses to quantify agreement and bias. Longitudinal consistency of wearable-derived physiological signals should also be evaluated in the context of known IBD flare physiology, where systemic inflammatory activity correlates with changes in inflammatory biomarkers and symptom indices [15]. Particular attention should be paid to signal drift, missing data patterns, and inter-device variability, as these factors directly influence digital twin calibration.

At the modeling level, validation should focus on the hybrid model's ability to infer latent colorectal states that are not directly observable. Inflammatory burden estimates derived from multimodal signals should be compared against serum C-reactive protein levels, fecal calprotectin, and validated clinical activity indices in longitudinal cohorts. Model-derived motility efficiency parameters may be evaluated indirectly through correlations with reported constipation, diarrhea frequency, or, where available, motility testing. For colorectal cancer risk modeling, predicted neoplastic transformation risk trajectories should be compared with colonoscopy findings, polyp detection rates, and histopathology outcomes over time.

Crucially, validation must also assess whether Colorectal Twin improves early detection of clinically meaningful events. For IBD, this includes prospective evaluation of flare prediction accuracy and time-to-flare detection relative to standard symptom-based care. For CRC, validation should assess whether longitudinal risk modeling enables earlier referral for colonoscopy or improves detection of advanced adenomas compared with age-based screening alone. For DD, evaluation may focus on detection of early inflammatory episodes prior to hospital admission.

Beyond predictive performance, implementation studies should examine whether digital twin outputs influence clinical workflow in a safe and interpretable manner. Metrics such as reduction in emergency admissions for IBD flares, optimization of surveillance colonoscopy intervals, decreased unnecessary procedures, and patient-reported quality-of-life improvements are essential indicators of translational value.

## POTENTIAL IMPACT

Colorectal Twin has the potential to transform colorectal care by enabling continuous, home-based monitoring and more proactive disease management. By reducing reliance on frequent clinic visits and repetitive laboratory testing, such a system could alleviate healthcare burdens while improving patient quality of life.

For clinicians, Colorectal Twin may serve as an advanced decision-support framework, offering deeper physiological insight and facilitating earlier, more personalized interventions. Rather than replacing established clinical pathways, the system is envisioned to augment existing care models by providing longitudinal, patient-specific context for diagnosis and treatment planning.

More broadly, this framework establishes a conceptual foundation for future digital twin applications across other organ systems and chronic diseases, supporting a shift toward continuous, data-driven, and personalized healthcare.

## ETHICAL, PRIVACY, AND CLINICAL INTEGRATION CONSIDERATIONS

The development of a colorectal digital twin requires robust governance mechanisms to ensure patient safety, privacy protection, and clinical accountability. Continuous multimodal monitoring introduces challenges related to data ownership, informed consent, and secure transmission of sensitive physiological and behavioral information. Privacy-preserving strategies such as edge processing, federated learning, encrypted data transmission, and role-based access control may mitigate risks while maintaining system performance.

Colorectal Twin is intended to augment—not replace—clinical decision-making. All high-risk outputs, including suspected disease progression or neoplastic transformation alerts, require clinician review before action. Integration with existing electronic health record systems should support bidirectional communication while preserving clinical workflow efficiency. These safeguards are essential for regulatory compliance and long-term translational viability.

## CONCLUSION AND FUTURE DIRECTIONS

This paper introduces Colorectal Twin as a conceptual digital twin framework for the early detection and personalized management of colorectal diseases. By integrating multimodal non-invasive data, physiologically grounded colorectal modeling, and adaptive artificial intelligence, Colorectal Twin seeks to address key limitations of current monitoring and management approaches.

Although this work does not present a fully implemented clinical system, it establishes a clear vision and defined approach for future methodological development, validation studies, and translational efforts. Realizing this vision will require interdisciplinary collaboration across gastroenterology, data science, biomedical engineering, and healthcare systems research.

Colorectal Twin represents a step toward continuous, patient-centered, and proactive colorectal care, with broader implications for extending digital twin methodologies to other chronic and organ-specific diseases. It is anticipated that early implementations will reveal practical integration constraints and modeling limitations that cannot be fully anticipated at the conceptual stage, necessitating iterative refinement.

## ACKNOWLEDGEMENT

ChatGPT was used to assist with language refinement, grammar correction, and structural clarity. All scientific content, interpretations, and conclusions were independently developed, validated, and approved by the team.